## RESEARCH ARTICLE

## IN VITRO ANTIBACTERIAL ACTIVITY OF HEXANE, CHLOROFORM AND METHANOLIC EXTRACTS OF DIFFERENT PARTS OF *ACRONYCHIA PEDUNCULATA* GROWN IN SRI LANKA.

*R.D. Nimantha Karunathilaka[1], Athige Rajith Niloshan Silva[2], Chathuranga Bharathee Ranaweera[3], D.M.R.K. Dissanayake[4], N.R.M. Nelumdeniya[5], Ranjith Pathirana[2] and W. D. Ratnasooriya[2].

1. Department of Nursing and Midwifery, Faculty of Allied Health Sciences, General Sir John Kotelawala Defence University, Werahere, Sri Lanka
2. Department of Basic Sciences, Faculty of Allied Health Sciences, General Sir John Kotelawala Defence University, Werahera, Sri Lanka.
3. Department of Medical Laboratory Sciences, Faculty of Allied Health Sciences, General Sir John Kotelawala Defence University, Werahere, Sri Lanka.
4. Department of Chemistry, University of Colombo, Colombo, Sri Lanka
5. Department of Pharmacy, Faculty of Allied Health Sciences, General Sir John Kotelawala Defence University, Werahera, Sri Lanka.

……………………………………………………………………………………………………....

*Abstract*
………………………………………………………………
This study accessed the antibacterial potential *in vitro* of hexane, chloroform and methanol extracts made from leaves, stem bark, flowers, seeds or roots of Sri Lankan grown *Acronychia pedunculata* plant against two Gram positive bacteria, *Staphylococus aureus* (ATCC 25923) and *Bacilus cereus* (ATCC 11778), and two Gram negative bacteria, *Pseudomonas aeruginosa* (ATCC 9027) and *Escherichia coli* (ATCC 35218), using agar disc diffusion bioassay technique. The results showed that none the of the extracts provoked an antibacterial action against the two Gram negative bacteria *P. aeruginosa* and *E. coli*. Conversely, compared to reference drug, Gentamicin, varying magnitudes of antibacterial activity (concentration: 300 mg/disc) ranging from zero to mild to moderate to strong antibacterial activity was evident with the three solvent systems made from different parts of the plant against the two Gram positive bacteria *S. aureus* and *B. cereus*. All the three flower extracts excerted marked antibacterial activity against both *S. aureus* and *B. cereus*. The highest antibacterial activity was exhibited by methanol flowers extract (inhibition zone: 13.8±0.32mm), with a Minimum inhibitory value of 32mg/ml, against *B. cereus*. The overall order of potency against *S. aureus* was, chloroform flowers> chloroform seeds > hexane leaves > chloroform leaves > methanol flowers> hexane flowers> methanol seeds. And against *B. cereus* was methanol flowers> hexane leaves > hexane flowers> chloroform leaves >chloroform flowers >chloroform seeds > hexane roots > chloroform roots > methanol seeds chloroform stem barks = hexane stem barks. These are all novel findings for *A. pedunculata* found in Sri Lanka and elsewhere. It is concluded that novel, potent and cost effective antibacterial agent/s, especially, against Gram-positive bacterial pathogens may be developed from flowers, seeds and leaves of *A. pedunculata*.

………………………………………………………………………………………………………....

**Corresponding Author:- R.D. Nimantha Karunathilaka.**
Address:- Department of Nursing and Midwifery, Faculty of Allied Health Sciences, General Sir John Kotelawala Defence University, Werahere, Sri Lanka.

………………………………………………………………………………………………………….....

## Introduction:-
*Acronychia pedunculata* L. (Family: Rutaceae) known as claw-flowered Laural and Lake Wood in English, Ankenda in Sinhala, and Kattukanni or Muttainari in Tamil, is a small tree with pale smooth bark and glabrous branches. The plant is generally found in parts of India, Malaysia, Philippine Islands and Sri Lanka (Jayaweera, 1982). In Sri Lanka, it grows in almost all parts of island except in extreme cold areas, in central hills above an elevation of 5000 feet (Jayaweera, 1982).

The leaves of this plant is oval in shape 7.5-12.5 cm in length, and arranged in simple, opposite or in alternate arrangement. Flowers are small, regular, polygamous pale yellowish green in colour, and arranged loosely in pyramidal or divaricately manner on long straight axillary peduncles. Flowering usually occurs between February and April. The fruits are indehiscent, typically globular shaped and about 1.2-1.8 cm long. Interestingly, fruits are four chambered and each chamber has one seed (Jayaweera, 1982).

In Sri Lanka traditional and folk medicine the leaves, stem bark and roots, in different forms/formulation, are used internally and cutaneously for swelling, inflammation, fractures, sores, scabies, skin rashes, boils (watery and pus filled) and for ulcers. [1] Stem bark is used as a purgative (Jayaweera, 1982; Rodrigo et al., 2007). In addition, the leaves and stem bark is used by villagers to prevent fermentation of toddy sap (Jayaweera, 1982). However, scientific validation and justification for these traditional uses are virtually none excitant or extremely limited.

Since its commonest use in Sri Lankan ethonomedicine appears to be in treatment of infective conditions in the skin we thought that, it may be possible to find novel antibacterial agent/s from different parts of this plant, including flower and seeds, which are not claimed to be used in treatment. Nevertheless, experimentally, we have shown that aqueous extracts of flower and seeds of this plant possess antibacterial activity *in vitro* (Ranaweera et al., 2016). Accordingly, in this study, we investigated the antibacterial potential *in vitro* of leaves, stem bark, roots, flowers and seeds of this plant using methanol, hexane and chloroform extracts and disc diffusion bioassay technique. These three organic solvents are widely used in the elucidation of antibacterial activity of botanicals (Ranaweera et al., 2015; Sastry and Rao, 1994; Silva et al., 2015) and so is the bioassay technique (Ranaweera et al., 2015; Sastry and Rao, 1994; Silva et al., 2015; Fancy et al., 1999).

## Materials and methods:-
### Collection and Authentication:-
Whole matured plants with flowers and fruits were collected in Mallawapitiya area in Kurungegala district in Sri Lanka (GPS 7°28'21.7"N 80°23'23.8"E), in April 2016. The whole plant had been identified and authenticated by a botanist at National Herbarium at Peradeniya, Sri Lanka.

### Preparation of hexane, chloroform and methanol crude extracts:-
The leaves, flowers, seeds, stem bark and roots were removed from the plants, washed thoroughly in tap water and air dried in shade for 3-4 days. Dried parts were cut into small pieces and were powdered using a mechanical grinder. Twenty grams of each of the powered samples were extracted into 200 ml of hexane, chloroform and methanol using sequential extraction techniques with sonification.

The extracted samples were filtered separately through double-layered muslin cloth and the filtrates were evaporated to dryness under reduced pressure. The resulting products were stored separately in air tight bottles at $4^0$C until use.

### Evaluations of the antibacterial activity:-
Antibacterial activity of hexane, chloroform and methanol extracts of different parts of the plant (disc concentration: 300 µg/disc) were evaluated using the agar disc diffusion method as described by Clinical and Laboratory Institute (NCCLS, 2003). Methanol was used as the negative control and Gentamycin (10 µg/disc) as the positive control. Two gram negative bacterial species, *Pseudomonas aeruginosa* (ATCC 9027) and *Escherichia coli* (ATCC 35218), and two gram positive species, *Staphylococcus aureus* (ATCC 25923) and *Bacillus cereus* (ATCC 11778) were used in the determination of the antibacterial activity. Incubation temperature was $37^0$C and incubation period was 24 hours. At the end of the incubation period the diameter of the transparent inhibition zone around each disc was measured (in mm) using a pair of vernire calipers. All experiments were conducted in triplicate. A sample was considered to have an effective antibacterial activity if the inhibition growth zone around the disc was equal or greater that 7.0 mm.

**Evaluation of Minimum Inhibitory Concentration (MIC):-**
The crude extracts that demonstrated effective antibacterial activities against *S. aureus* and *B. cereus* were subjected to the determination of MIC using broth micro-diffusion method with slight modification using Mueller Hinton broth as the medium as described by National Committee for Clinical Laboratory Standards (NCCLS, 2002). The MIC end point was considered as the lowest concentration with more than 90 % growth inhibition. The optical density of the bacterial growth was determined (at 600 nm) using a DTX 880 plate reader. The commercial antimicrobial agents Agentsopolymixin B and Rifamycin were respectively used as positive controls for *B. cereus*, for *S. aureus*. Concentration series used was 2.0-0.004 µg/ml.

## Statistical Analysis:-
Data is expressed as descriptive statistics (mean ± SEM) using SPSS (17$^{th}$ Version).

## Results:-
The results are summarized and depicted in Tables 1 and 2.

**Table 1:-** *In vitro* antibacterial activity of plant extracts of *Acronychia pedunculata* against *Staphylococcus aureus* and *Bacilus cereus*

| Parts of *Acronychiapedunculata* plant | Extract Type | Anti-microbial activity (300 µg/disc) (Mean inhibition zone diameter ± Standard error - mm) | |
|---|---|---|---|
| | | *Staphylococus aureus* | *Bacilus cereus* |
| Leaf | Hexane | 12.0±0.3 | 13.0±0.2 |
| | Chloroform | 10.3±1.6 | 10.0±0.2 |
| | Methanol | - | - |
| Bark | Hexane | - | 7.4±0.4 |
| | Chloroform | - | 7.3±0.4 |
| | Methanol | - | - |
| Root | Hexane | - | 8.4±0.2 |
| | Chloroform | - | 8.1±0.2 |
| | Methanol | - | - |
| Flower | Hexane | 9.0±0.6 | 12.5±1.3 |
| | Chloroform | 13.1±0.1 | 9.0±0.4 |
| | Methanol | 9.3±0.8 | 13.8±0.3 |
| Seed | Hexane | - | - |
| | Chloroform | 12.8±0.5 | 8.6±0.1 |
| | Methanol | 7.2±0.5 | 7.6±0.7 |
| +Ve control | | 16.1 ±1.0 | 18.6 ±0.8 |
| -Ve control | | - | - |

Sign (-) indicates no antibacterial activity.
None of the extracts made from different parts of *A. pedunculata* at 300 mg/disc concentration, had any antibacterial activity (in terms of the diameter zone of inhibition) against Gram-negative bacteria, *P. aeruriginosa* and *E. coli*. Similarity, methanol, leaves, stem bark, roots, and hexane seed extracts were ineffective against the two Gram positive bacteria, *S. aureus* and *B. cereus*. Further, hexane and chloroform extracts of stem bark and roots were ineffective against *S. aureus* but not against *B. cereus* (See Table 01). On the other hand, hexane, chloroform and methanol extracts of flowers, hexane and chloroform extracts of leaves, and chloroform and methanol extracts of seeds induced marked and significant (P<0.05) antibacterial activity against both *S. aureus* and *B. cereus*. In complete contrast, hexane and chloroform extracts of stem bark and roots had significant (P<0.05) antibacterial activity only against *B. cereus*. Rifamycin and Polymyxin B provoked strong antibacterial actions respectively against *S. aureus* and *B. cereus*.

MIC values of different extracts against *S. aureus* and *B. cereus* is given in Table 2.

**Table 2:-** MIC values of different extracts *Acronychia pedunculata* against *S. aureus* and *B. cereus* is

|  |  | Anti-microbial activity (concentrations are in µgmL$^{-1}$) | |
|---|---|---|---|
| Parts of *Acronychia pedunculata* plant | Extract Type | *Staphylococus aureus* | *Bacilus cereus* |
| Leaf | Hexane | 32 | 32 |
|  | Chloroform | 64 | 64 |
|  | Methanol | - | - |
| Bark | Hexane | - | - |
|  | Chloroform | - | - |
|  | Methanol | - | - |
| Root | Hexane | - | 256 |
|  | Chloroform | - | 256 |
|  | Methanol | - | - |
| Flower | Hexane | 128 | 256 |
|  | Chloroform | 32 | 128 |
|  | Methanol | 128 | 32 |
| Seed | Hexane | - | - |
|  | Chloroform | 32 | 32 |
|  | Methanol | - | - |
| +Ve control |  | 0.125 | 2 |
| -Ve control |  | - | - |

Sign (-) indicates no MIC value.

As shown, the lowest MIC values of 32 µg/ml against *S. aureus* was seen with hexane leaves and chloroform flower extracts whilst the same MIC values were evident for hexane leaves and chloroform seeds extracts against *B. cereus*. The highest MIC value of 256 µg/ml were evident with hexane flower extract and hexane and chloroform roots extracts. MIC values for Rifamycin against *S. aureus* and Polymyxin B against *B. cereus* were remarkable low, respectively, 0.125µg/ml and 2.0 µg/ml.

## Discussion:-

Using *in vitro* studies we have previously shown that hot water extracts of seeds, flowers, leaves, stem bark and combination of these parts of *A. pedunculata* have mild to moderate antibacterial activity against Gram positive bacteria, *S. aureus* but not against Gram negative bacteria, *E. coli* (Ranaweera et al., 2016). In this study, we investigated *in vitro* antibacterial activity of hexane, chloroform and methanol extracts of leaves, flowers, seeds, and roots of *A. pedunculata* against two Gram positive pathogenic bacterial species, *S. aureus* and *B. cereus*, and two Gram negative pathogenic bacterial species, *P. aeruriginosa* and *E. coli,* using agar disc diffusion bioassay. Both bioassay model and the solvents (hexane, chloroform and methanol) used for extraction are commonly used in the evaluation of antibacterial potential of natural products Ranaweera et al., 2015; Sastry and Rao, 1994; Silva et al., 2015; Fancy et al., 1999).

The results explicitly showed that under our experimental conditions, none of the extracts was effective against the two Gram negative bacteria, *P. aeruriginosa* and *E. coli*. Absences of any antibacterial activity against Gram-negative bacteria in this study may be a general phenomenon evident with herbal antibacterial agents or even with the therapeutically used antibiotics (Wang et al., 2002; Hamilton – Millar, 1995; WHO, 2016). Cell wall of Gram-negative bacteria can suppress the influx of antibacterial agents into the cytosol (WHO, 2016). Alternatively, it may be due to absence of hydrophobic and hydrophilic phytoconstituents in *A. pedunculata* found in Sri Lanka which could exert an antibacterial action against Gram negative bacteria: hexane and chloroform are none polar solvents and methanol is a polar solvent.

On the other hand, varing degree of antibacterial activity was evident against both Gram positive bacteria, *S. aureus* and *B. cereus* the depending on the plant part tested and the type of solvent used in the extraction process: methanol extracts of leaf, stem bark, roots and hexane extract of seed exhibited no antibacterial activity against both *S. aureus* and *B. cereus* whilst hexane and chloroform of stem bark and roots were effective against *S. aureus*.

In complete contrast, all hexane, chloroform and methanol leaf extracts and chloroform and methanol extracts of seed induced mild to strong antibacterial activity against both *S. aureus* and *B. cereus*. Importantly, the highest activity against *S. aureus* and *B. cereus* were seen respectively with chloroform flower extracts and methanol flower extract. The overall order of potency against *S. aureus* was, chloroform flowers > chloroform seeds > hexane leaves > chloroform leaves > methanol flowers > hexane flowers > methanol seeds. And against *B. cereus* was methanol flowers > hexane leaves > hexane flowers > chloroform leaves > chloroform flowers > chloroform seeds > hexane roots > chloroform roots > methanol seeds chloroform stem barks = hexane stem barks.

On the other hand, the order of potency reported for antibacterial activity of different parts of aqueous extract of aqueous extract of Sri Lankan grown *A. pedunculata* against *S. aureus* was combined extract > roots > stem bark > leaves > flowers > seeds (Ranaweera et al., 2016). However, in agreement with this study, none of the aqueous extracts of *A. pedunculata* was effective against Gram negative, *E. coli* (Ranaweera et al., 2016). Collectively, this study together with our pervious study (Ranaweera et al., 2016) suggest firstly that different parts of Sri Lankan grown *A. pedunculata* plant do not contain hydrophobic and /or hydrophilic phytoconstituent which has antibacterial activity against Gram negative bacteria, at least, against *E. coli*. Secondly our pervious (Ranaweera et al., 2016) and this study suggest that leaves, flowers and seeds of *A. pedunculata* possess both hydrophilic and hydrophobic phytoconstituent which have antibacterial action against Gram positive, *S. aureus* whilst roots and stem bark have only hydrophilic phytoconstituent exhibiting such activity. On the contrary, leaf, stem bark, roots, flower and seed extracts of *A. pedunculata* contained hydrophobic phytoconstituent which have antibacterial action against Gram positive bacteria, namely *B. cereus*. Unfortunately, at present, we do not know whether this plant contain any hydrophilic phytoconstituent which can confer antibacterial action against Gram positive, *B. cereus*, as in our previous study, we have not accessed antibacterial activity against this bacterial species (Ranaweera et al., 2016).

MIC values obtained in this study show that hexane leaf extract, chloroform flower extract and chloroform seed extract has the strongest and highest activity against *B. cereus* whilst hexane leaf extract and chloroform seed extract has the strongest and highest activity against *B. cereus*.

### Conclusion:-
In conclusion, the results of this study, showed the first time that, hexane, chloroform and methanol extracts of different parts of Sri Lankan grown *A. pedunculata* possess varing degree (mild to strong) of antibacterial activity against gram-positive *S. aureus* and *B. cereus*. To our knowledge, no studies are reported on antibacterial activity of *A. pedunculata* grown elsewhere. The results also indicate that different parts of Sri Lankan grown *A. pedunculata*, especially, the flower, seed and leaves offer high promise to develop novel, safe and cost effective antibacterial agents. However, further studies on isolation and characterization are mandatory in this regard.

### Acknowledgement:-
Thanks are due to Mr H.A.K.I Dharmakeerthi for collection of the plants and Dr. A.K Chandana for assistance in collecting reference material.

### References:-